\documentclass[a4paper, 12pt]{article}
\usepackage[margin=1.2in]{geometry}
\usepackage{amsmath,amsfonts,amsthm} 
\usepackage{graphicx,cite,times}
\usepackage{enumitem,version}
\usepackage{pdfpages,url}
\usepackage{setspace}
\numberwithin{equation}{section}
\RequirePackage[colorlinks,citecolor=blue,urlcolor=blue]{hyperref}
\newcommand{\orcid}[1]{\href{https://orcid.org/#1}{\textcolor[HTML]{A6CE39}{\aiOrcid}}}
\Large

\begin{document}
	\title{A review on mathematical strength and analysis of Enigma}
	\author{Kalika Prasad\footnote{E-mail: klkaprsd@gmail.com, ~~ORCiD: ~\url{https://orcid.org/0000-0002-3653-5854/}} ~and Munesh Kumari\footnote{E-mail: muneshnasir94@gmail.com}
	\\\normalsize{$^{1,2}$Department of Mathematics, Central University of Jharkhand, India, 835205}}      
	\date{\today}
	\maketitle
	\noindent\rule{16cm}{.1pt}
	\begin{abstract}
		In this review article, we discussed the Mathematics and mechanics behind the Enigma machine with analysis of security strength.		
		The German army used the Enigma machine during the second world war to encrypt communications. Due to its complexity, the encryption done by the Enigma Machine was assumed to be almost unbreakable. However the Polish believed that people with a good background and deep knowledge of science and mathematics would have a better chance to break the encryption done by Enigma. They appointed twenty mathematicians from Poznan University to work on this problem at Polish Cipher Bureau. Three of those, Marian Rejewski, Jerzy Rozycki and Henryk Zygalski were able to exploit certain flaws in the encryption, and by using permutation group theory finally managed to decipher the Enigma messages. The mathematics discovered by them is presented here.
	\end{abstract}
	\noindent\rule{16cm}{.1pt}
	\section{Introduction}
	 
	\subsection{The Enigma Machine}
		Enigma is derived from the Latin word aenigima, which means mysterious, Cipher is from the Arabic word sifr (empty). In cryptography\cite{stallings2017cryptography} this word is used for the algorithm that does encryption or decryption. 
		
		Enigma Cipher Machine\cite{miller1995cryptographic,enigmaweb,EnigmaMachine,accordingtobenedict} is the name of an electro-mechanical rotor cipher machine which was first developed in Germany to provide security to commercial communication among companies and banks just after the 1st world war, around 1917.
		The Dutch invented these rotor based cipher machines earlier in 1915, however  officially the Enigma Cipher Machine was first developed by Arthur Scherbius in 1918 and Enigma was just a brand name. Later several versions of Enigma were developed and named Enigma A, B, C, etc.
		\begin{figure}[h]
			\centering
			\includegraphics[width=5.34in,height=7.6cm]{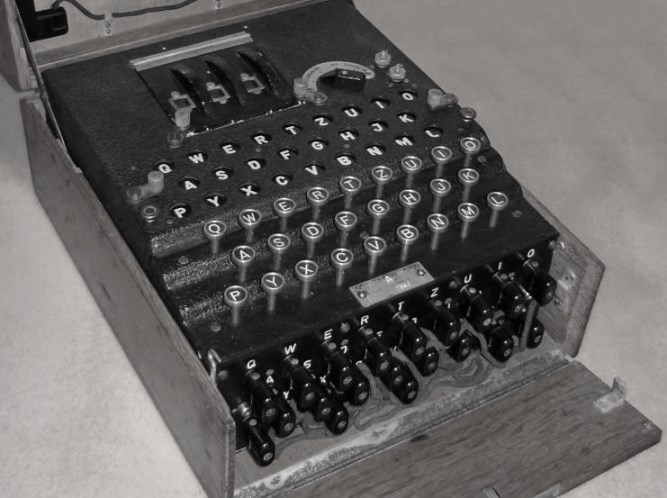}
			\caption{Exterior view of Enigma Showing front plugboard with cables\protect\footnotemark}
		\end{figure}
		\footnotetext{Source:  \url{http://users.telenet.be/d.rijmenants/en/enigmatech.htm}}
		
		In 1926 the German Army started taking interest in Enigma machines due to its specific and high security strength. In 1927 they build their first prototype, and used it exclusively to test its capacity. The final version for use in the German army was Enigma-I and was ready for deployment by 1932.
		Initially the Enigma cipher machines were free for use by any organisation. But after 1932, the German army changed the rule and there after use of any Enigma machine had to be approved by the German Army. By this time many other countries were also involved in making such cipher machines.\\
		\\In 1930, the Polish Cipher Bureau started taking interest in decoding the messages sent using Enigma. They started doing research on the commercial Enigma machines initially. The German army cipher machine appeared as big challenge at that time. The Polish bureau believed that mathematicians and scientists would have high chance of breaking the Enigma cipher. So the Cipher Bureau of Poland appointed 20 mathematicians from Poznan University to work on the Enigma machines. Marian Rejewski, Jerzy Rozycki and Henryk Zygalski were three students who achieved a breakthrough in decrypting the Enigma Cipher. In fact, Marian Rejewski achieved his first success within week of his appointment in 1932.
		\begin{figure}[h]
			\centering
			\includegraphics[width=5.34in,height=6cm]{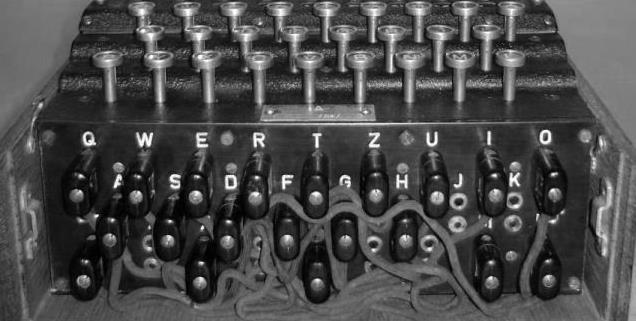}
			\caption{Close-up of Plugboard with cables \protect\footnotemark}
		\end{figure}
		\footnotetext{Source:  \url{http://users.telenet.be/d.rijmenants /en/enigmatech.htm}}
		
		Here, we discuss the study report of structure of the Enigma Machine and then analyse the strength of machine using Mathematics. This paper is the extension of same work on 'the cryptographic mathematics of enigma'\cite{miller1995cryptographic} to analyze the security and strength of machine. 
		
	\section{Structure}
	\begin{figure}[h]
		\includegraphics[height=5.6cm,width=\textwidth]{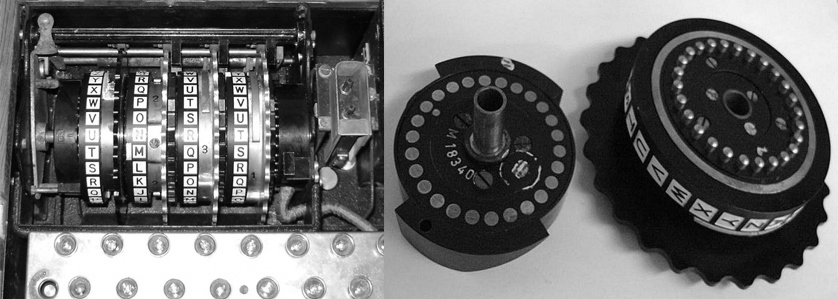}
		\centering
		\caption{Overview of Rotors\protect\footnotemark}
	\end{figure}
	\footnotetext{Source: (a). \url{http://users.telenet.be/d.rijmenants/pics/hires-m4inside.jpg}			
		(b). \url{http://users.telenet.be/d.rijmenants/pics/hires-m4ukwbeta.jpg}}\vspace{1cm}	
	
	There were many variants of the Enigma; the most common version was the standard Army and Luftwaffe Enigma machine. Important components of Enigma were
		\subitem (i)\hspace{.3cm} A Plugboard;
		\subitem (ii) \hspace{.1cm} A set of five rotors;
		\subitem(ii) \hspace{.1cm} A Keyboard;
		\subitem(iv) \hspace{.07cm} A Lampboard;
		\subitem (v) \hspace{.2cm} Reflectors.
		
		\begin{figure}[h]
			\includegraphics[height=5cm,width=\textwidth]{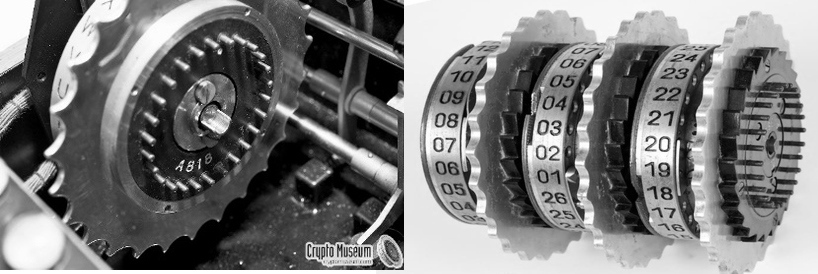}
			\caption{Reflector and Rotor\protect\footnotemark}
		\end{figure}
		\footnotetext{Source: (a). \url{http://www.cryptomuseum.com/crypto/enigma/} \\(b).\url{http://www.cryptomuseum.com/crypto/enigma/i/img/300002/050/full.jpg}}

	\section{Further Description of the Components}
	Important variables components of an Enigma ciphering machine are as follow \cite{miller1995cryptographic}
	\begin{enumerate}
		\item The plugboard, which contained dual-wired cables from zero to thirteen.
		\item The set of five rotors had three rotors that were ordered(left to right) in such a way that 26 input points to 26 output points were positioned on alternate faces of the discs. The First rotor was a stecker (static wheel) and the last (end) rotor was called a Reflector. In some Enigma machines, they used three rotors and two were optional.
		\item There are 26 serrations around the periphery of the rotors which allowed the operators to specify an initial rotational position for the rotors,
		\item To control the rotational behavior of the rotor, there was a movable ring on each of the rotors placed immediately to the left by means of a notch.
		\item At the end there was a reflector(a half-rotor fixed at the end), to return inputs and outputs back onto the same face of contact points.
	\end{enumerate}
	\pagebreak
	Following figure shows the structure and the schematics of the Enigma cipher machine.
		\begin{figure}[h]
			\includegraphics[width=5.6in]{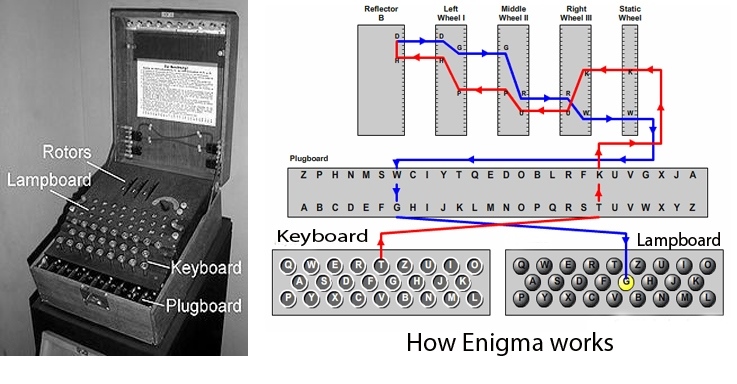}
			\caption{Assembled machine with all components\protect\footnotemark}
		\end{figure}
		\footnotetext{Source: (a). \url{https://en.wikipedia.org/wiki/Enigma-machine} \\(b). \url{http://enigma.louisedade.co.uk/howitworks.html}}
		
	\section{Functioning of Machine and Mathematics}
		Now, here we wish to determine how many number of different ways of configuring the variable components in the system which contributed to the cryptographic strength of the machine.
	\subsection{First variable components : Plugboard}
		The front panel of Enigma contains $26$ dual-holed$(A-Z)$ sockets on the front panel of Enigma. For making connection between any pair of letters, 
		a dual-wired plugboard cable were inserted.
	\subsection*{Total number of possible plugboard connections}
		To calculate the total number of possible plugboard connection, consider the following three components which played important roles:
		\subitem (a). The total number of cables used.
		\subitem (b). Group  of the sockets(which were there to receive those cables used) and
		\subitem (c). Total interconnections within one group to other group of sockets (i.e setting of specific letter-pairs made by individual cables).
		\begin{figure}[h]
			\includegraphics[height=8cm,width=\textwidth]{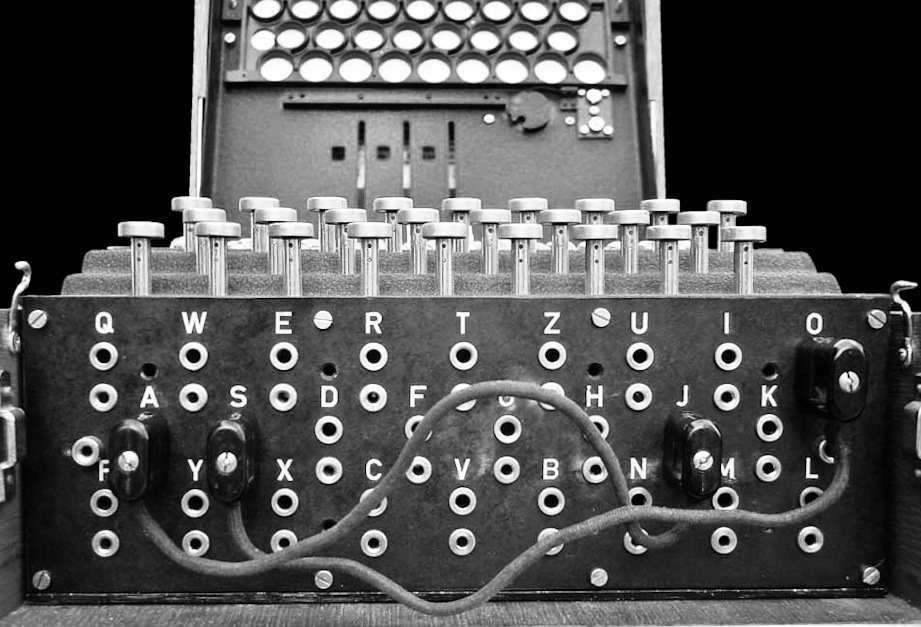}
			\caption{Plugboard sockets with and without cable.\protect\footnotemark}
		\end{figure}
		\footnotetext{Source: \url{https://commons.wikimedia.org/wiki/File:Enigma-plugboard.jpg}}
		\\Further Calculation,
	\subsubsection{The number of cables used}
		First, We will consider sockets selection. Since there are 26 sockets on plugboard and each individual cables have two ends, so each cable used two sockets(One for each end).
		\\Since, we have $p$ plugboard cables to connect with the plugboard	and one cable uses two sockets and we have 26 sockets in total. Therefore we have total 13 cables i.e $0 \leq p\leq 13$.
	\subsubsection{Sockets selection}
		Since we have to choose cables from those 26-cables, so total possible different combinations of sockets are  $(^{26}C_{2p})$ or $\dbinom{26}{2p}$.
		\\ Here we will determine in how many different ways these $p$ cables can be connected with those $2p$ selected sockets.
		\\ First, insert the first end of first cable into any of those $2p$ socket and then the second end of the first cable, which has total $(2p-1)$ free sockets available to choose.
		\\Next, insert the first end of second cable and thus the second end of second cable has $(2p-3)$ free sockets available to choose.
		\\This pattern continuous down up to last one cable, whose second end has only one free socket left to connect.
		\\Thus, the total possible number of ways in which those $p$-cables could have been connected with those $2p$ sockets are 
		\begin{eqnarray}
			(2p-1)(2p-3)(2p-5)...(3)(1) = (2p-1)!!
		\end{eqnarray}
		Therefore, the number of different connection which could have been made by an Enigma cipher machine is given by\cite{miller1995cryptographic}
		\begin{eqnarray}\label{TotalConn}
			^{26}C_{2p} \times (2p-1)!!=\dfrac{26!}{(26-2p)!\times p! \times2^p}
		\end{eqnarray}		
		\begin{proof}
			As L.H.S.
		\begin{eqnarray}
				^{26}C_{2p}\times ((2p-1)!!) &=& \dfrac{26!}{(26-2p)!\times2p!}\times(2p-1)(2p-3)...(3)(1) \nonumber
				\\&=&\dfrac{26!}{(26-2p)!\times(2p)(2p-2)(2p-4)....(4)(2)}\nonumber
				\\&=&\dfrac{26!}{(26-2p)!\times(2.p).2.(p-1).2.(p-2)...(2.2).(2.1) }\nonumber
				\\&=&\dfrac{26!}{(2p-2p)!\times 2^p.(p-1).(p-2)....2.1}\nonumber
				\\&=& \dfrac{26!}{(26-2p)!\times p! \times2^p} \nonumber
				\\&=& R.H.S \nonumber
		\end{eqnarray}
		\end{proof}
				
	\subsubsection{Possible plugboard Combination }
		As, it has been shown above that $^{26}C_{2p} \times (2p-1)!!=\dfrac{26!}{(26-2p)!\times p! \times2^p}$. So, from the relation(\ref{TotalConn}), the number of all possible plugboard combination for all value of p are as follows:
		\begin{eqnarray}\label{table1}
			\begin{tabular}{|c|c||c|c|}
				\hline
				Values of p & Combinations & Values of p & Combinations\\[0.5ex]
				\hline \hline
				0 & 1 & 7 & 1,305,093,289,500\\
				\hline
				1 & 325 & 8 & 10,767,019,638,375 \\
				\hline
				2 & 44850 & 9 & 53,835,098,191,875\\
				\hline
				3 & 3,453,450 & 10 & 150,738,274,937,250\\
				\hline
				4 & 164,038,875 & 11 & 205,552,193,096,250\\
				\hline
				5 & 5,019,589,575 & 12 & 102,776,096,548,125\\
				\hline
				6 & 100,391,791,500 & 13 & 7,905,853,580,625\\
				\hline
			\end{tabular}
		\end{eqnarray}
		Here, there is a interesting observation from above table that the maximum number of combinations did not occurs at 13 as we might expect; but it occurs at $p=11$.
		\\Since the combinations made up here for each value of p are mutually exclusive, so the total number of possible plugboard combination is given by summing all the above cases, i.e.
		\begin{eqnarray}\label{PlugBcombination}
			\sum_{p=0}^{13}\dfrac{26!}{(26-2p)!\times p! \times2^p } = 532,985,208,200,576
		\end{eqnarray}
	\subsection{The second variable component}
		\begin{figure}[h]
			\centering
			\includegraphics[scale=0.4]{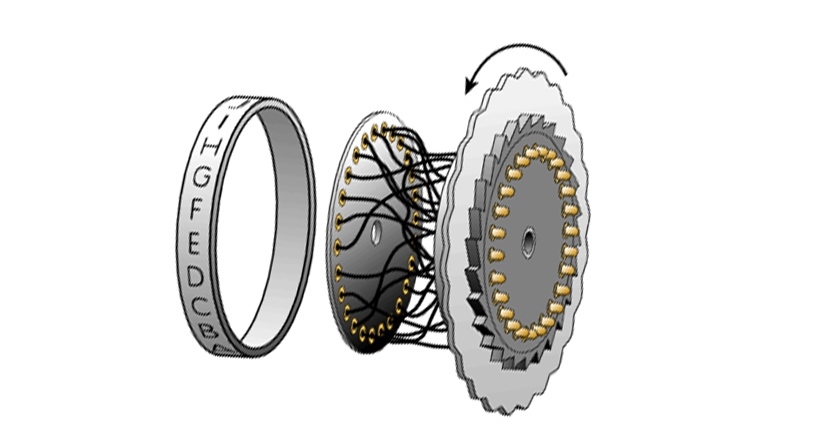}
			\caption{Disassembled Rotor with 26 input and output point on alternating faces\protect\footnotemark}
		\end{figure}
		\footnotetext{Source: \url{http://www.cryptomuseum.com/crypto/enigma/}}
		The second variable component was the set three ordered rotors which connects to those $26$-input contact points to $26$-output contact points positioned on both sides of a disc. So, total number of discs which could have been constructed independently are $26!$.
		\\ Note that:
		\begin{eqnarray}
			26! = 403,291,461,126,605,635,584,000,000
		\end{eqnarray}
		Since the rotor's discs are hardwired. So, such a vast number would have been impossible in practice to construct.
		\\(Also, The Germans never changed the disc wirings during the war. They did, however create several different groups of rotor's disc wirings for  special purpose machine.)
		\\So to place the leftmost position of rotor, the cryptographers could have select any one of disc from those $26!$.
		Further, for the middle position, it could have been selected from any one of the $(26!-1)$ discs and for the rightmost, it could have been selected from any one of $(26!-2)$ discs.
		\\So, the total number of ways of all possible combination of ordering in the Enigma machine is
		\begin{eqnarray}\label{OrderComb}
			26!\times (26!-1) \times (26!-2)
		\end{eqnarray}
		\textbf{Remark :}
		\\It was known that the German troops carried individually numbered and unique sets of rotors. Hence, selection of a rotor reduced the no. of possibilities by one. So $26!^3$ is not the correct value.
		\subsection{The third variable component : Initial rotational position}
		Initial rotational position of the three rotors containing the wired discs was the important variable component.
		This was specified and set by cryptographers(also by machines operators). Since, It had 26 serrations around the rotor periphery and there were three rotors with each has connected to one of the 26 different positions. So, the total number of key setting(combinations of rotor) was $26^{3}=17,576$.
		
		\subsection{The fourth variable component : Movable ring}
		As we already discussed, there were movable ring on each of the rotors. Further,
		each ring contained a notch\footnote{the notch purpose was to force a rotation of the rotor immediately to the left when the notch was in particular position}. Thus, when a key was pressed the right most rotor rotated every time(as the notch of rightmost rotor forced a rotation of the middle rotor once every 26 keystrokes). Similarly, for middle rotor, it's notch forced a rotation of the leftmost rotor once every $26\times 26$ keystrokes. Since there are no more rotors there so the leftmost rotor has no effect. So the total possible combination is $26^{2}= 676$.
		
		\subsection{The fifth variable component : The Reflector}
		As we know the reflector  was placed at the end and it had 26 contact points like a rotor, but only on one face. On that one face of reflector, 13 wire internally connected to the 26 contact points together in a series of pairs so that a connection input on the reflector from the rotors could fold back through the rotors but by the a different route.
		
		\subsubsection*{Internal Wiring}
		Start connecting of wires one-by-one, take one end of first wire and connect it with any one of the contact points, thus for the other end of the wire has total $25$ different choices(contact points) to connect, now connect it. Thus the first wire used two contact points with 25 different possibilities. Next, in similar way, second wire also consume two of contact points, and it has only 23 different possibilities. For rest wires, continue in similar way till the last contact point consumed.\\
		So, the total number of distinct reflector connection which could have been placed into enigma are given by
		\begin{eqnarray}\label{IntWiring}
			25!! = \dfrac{26!}{(13!\times 2^{13})}= 7,905,853,580,625.
		\end{eqnarray}
		\begin{figure}[h]
			\centering
			\includegraphics[height=7cm,width=\textwidth]{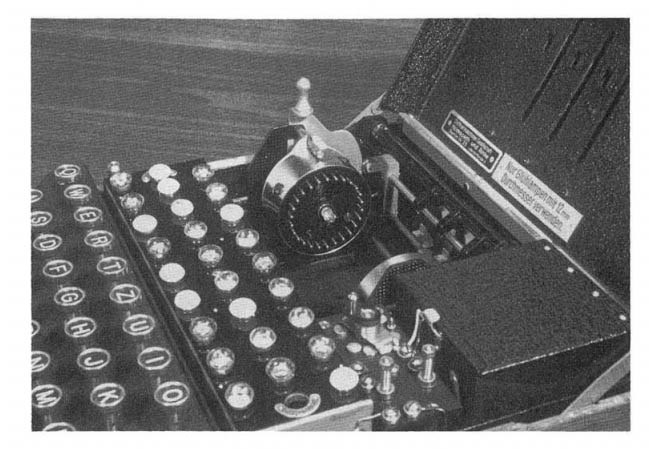}
			\caption{Reflector with rotors removed\protect\footnotemark}
		\end{figure}
		\footnotetext{Source: \url{https://en.wikipedia.org/wiki/Enigma-machine}}
	     Here interesting fact is that the total number of different reflector combinations is also same as the total number of possible plugboard combinations when $13(=p)$ cable were used(\ref{table1}).
	\section{Theoretically total possible Enigma configuration}
		Here, the total theoretical number of possible  Enigma configuration is simply the product of all five variable components which is approximately $3\times 10^{114}$.
		\\You can imagine how large this number is, as it is estimated that there are only about $10^{80}$ atoms in the entire observable universe. No  wonder, the German Cryptographer had confidence in their machine!
		
		German Forces used to the most common model 3-rotor and single notched Enigma. However later in the war, the German navy adopted a variant version of the Enigma Cipher machine which used four rotors \& rings and also it contained either a single or dual notch.\\
		\\To check how these modification increased security of machine, let us recalculate the theoretical number of Key settings and machine configuration for  a natural Enigma. For that purpose, we follow the following steps :-
		
	\subsubsection*{Step-I  : Total number of plugboard combinations}
		Here in this case, one extra rotor and notches added to the Enigma machine which has no effect on the plugboard combination. So total plugboard combination will be unchanged i.e $532,985,208,200,576$ (See, \ref{PlugBcombination})
		
	\subsubsection*{Step-II : Selection and Ordering of wired Rotor discs}
		As previously calculated(\ref{OrderComb}), It was $26!(26!-1)(26!-2)$. So for one extra rotor (i.e. fourth rotor), simply add$(26!-3)$ in the factor.
		Since, the new fourth rotor was not interchangeable with the other rotors; it could only be placed on one location(positioned on the leftmost next to the reflector).
		This means the $4^{th}$ rotor was incompatible, and could not be used in the place of other three rotors. So the selection of the $4^{th}$ rotor was independent, and thus for it, there were 26! possibilities.
		\\Thus, total possibilities in this case is given by $26!\times(26!-1)\times (26!-2)\times 26!$
		
	\subsubsection*{Step-III : Initial Rotational position of the wired discs}
		Since in the new enigma machine, one extra rotor had been added, which had 26 wired discs. So total possible combinations are $26^{4}$.
	\subsubsection*{Step-IV : Initial positions of movable notch rings}
		To increase the irregularity of the rotational behavior of the rotors, the German Navy used a second notch to some rings in order.
	\subsubsection*{Total possible combination of single or dual notched rings}
		Since, for a ring with a single notch had 26 possible orientations and with dual notch had $26\times25$ possible orientations and both the cases were mutually exclusive. So for this case, total number of  possible combinations
		\begin{eqnarray}
			&=& 26+26\times25 \nonumber\\
			&=& 26^{2}
		\end{eqnarray}
		Now, on three rotor or four rotor Enigmas(four rotor Naval Enigmas), the notch location only mattered for the rotors placed into the middle position and  rightmost and does not on leftmost.
		Since the 4th rotor has no ratchets and the Enigma has no 4th steeping lever, the 4th rotor did not move. Once the Enigma has  set the initial rotational position by hand, it remained constant for the duration of message. So the total number of possible single or double notched ring positions is $26^{4} = 456,976$.
	
	\subsubsection*{Step-V : Reflector Wiring}
		The total number of possible wiring configuration does not change as we calculated in(\ref{IntWiring}) i.e total variable component are $7,905,853,580,625$.
		Because the Navy introduced just one half width reflector so that the machine could continue to fit the same size space.
		
		Now from above all calculations, we are ready to determine the total theoretical number of possible Naval Enigma's configuration(assuming that,  4-rotors and single notches in the Rings).
		As we previously calculated total possibilities, similarly here it is the product of all the five values calculated above, which is approximately 2$\times 10^{145}$.
		
		The numbers derived thus far are theoretical values which reflect that initially how many cryptographic settings were possible.

	\section{Possible Crypt-variables}
		Finally, here we will calculate the total number of possible crypt-variables. 
	\subsubsection*{In Step-I :  The plugboard}
		In the plugboard, the most value of $p$ used was $10$.
	And according to the table(\ref{TotalConn}) it is 150,738,274,937,250.
	\subsubsection*{In Step-II : The selection and ordering of rotor discs }
		"Initially only 3 rotor discs were created for general purpose use(Special purpose, as previously stated, had there own set of wirings). Later, two additional rotor discs were introduced making five total. The German Navy added an additional three rotors discs bringing their total to eight. And finally, one and then two extra fourth rotor discs (without rotation ratchets) were added by the Navy giving them 10 possible discs." as mentioned in \cite{miller1995cryptographic}.
		\\ Here, we will assume that this is an Enigma machine with 3 rotors and the general purpose case of five discs with wiring of each of the discs is known.  Here, target of Allied cryptanalysts was to determine that which 3 of the 5 possible discs were chosen and what is their order in the machine. So to find out correct settings, they need to check simply $\dbinom{5}{3}$ or $5\times4\times3=60$ possible combinations.
	\subsubsection*{In Step-III : The initial rotational position}
		Since, for key setting initial rotational position of the rotors was an unknown. So in this case total possibilities are $26^{3}$ or 17,576.
	\subsubsection*{In Step-IV : The position of the notched rings}
		Since, Dual notched rings were not introduced until the Navy added their extra three rotor discs. So, we will consider single notches on all of the rings.
		Therefore, the total number of positions of the notched ring are $26^{2}$ or 676.
	\subsubsection*{In Step-V:}
		As, in above Enigma machine, single reflector were used. And it's wiring is already known, so the number of possible combinations is 1.\\
		
		"Thus the possible cryptvariables space Allied cryptanalysts were typically faced with during the second world war when attempting to read Enigma traffic is the product of the above five values which is approximately $1\times 10^{23}$" according to\cite{miller1995cryptographic}. Observe that this value is much smaller than the total number of atoms in the entire observable universe. 
		\\"This is all the more true considering Allied cryptanalysts were faced with continually changing message keys on at least a daily basis - for every different radio network the Germans constructed".

	\section*{Acknowledgment}
		First author\footnote{First author is supported by DST-INSPIRE Scholarship 4406/2012} acknowledge to DST(Department of Science and Technology, India) for the financial support. Also thanks to Prof. Geetha Venkataraman for  her fruitful  guidelines \& discussion. We also thank Mitul Verma (Asst. Prof., Ashoka University, Haryana) for his cooperation and discussion.
	\bibliography{Enigma}
	\bibliographystyle{acm}
\end{document}